\documentclass[runningheads]{llncs}
\usepackage{latexsym}
\usepackage{bbm}
\usepackage[dvipsnames]{xcolor}
\usepackage{multirow}
\usepackage{url}
\usepackage{amsmath}
\usepackage{amsfonts,amssymb}
\usepackage{graphicx}
\usepackage{xspace}
\usepackage{booktabs}
\usepackage{adjustbox}
\usepackage{subcaption}
\usepackage{enumitem}
\usepackage{hyperref}
\usepackage{wrapfig}
\usepackage{caption} 
\setlist[itemize]{leftmargin=*}  

\newcommand{\CLEAR}{\textsc{clear}\xspace}
\newcommand{\ssl}[1]{{#1}$^\downarrow$}

\newcommand{\cmu}{\textsuperscript{\rm 1}}
\newcommand{\jhu}{\textsuperscript{\rm 2}}

\newcommand{\qry}{\langle\textsc{qry}\rangle}
\newcommand{\doc}{\langle\textsc{doc}\rangle}

\title{Complement Lexical Retrieval Model with \\ Semantic Residual Embeddings}

\author{
Luyu Gao\cmu, Zhuyun Dai\cmu, Tongfei Chen\jhu, Zhen Fan\cmu \\ 
Benjamin Van Durme\jhu, Jamie Callan\cmu
}
\authorrunning{L. Gao et al.}

\institute{\cmu~Carnegie Mellon University, \jhu~Johns Hopkins University}

\begin{document}

\maketitle
\begin{abstract}
This paper presents \CLEAR, a retrieval model that seeks to complement classical lexical exact-match models such as BM25 with semantic matching signals from a neural embedding matching model. 
\CLEAR explicitly trains the neural embedding to encode language structures and semantics that lexical retrieval fails to capture with a novel residual-based embedding learning method. Empirical evaluations demonstrate the advantages of \CLEAR over state-of-the-art retrieval models, and that it can substantially improve the end-to-end accuracy and efficiency of reranking pipelines. 
\end{abstract}
\section{Introduction}






 

State-of-the-art search engines adopt a multi-stage retrieval pipeline system: an efficient first-stage \emph{retriever} uses a query to fetch a set of documents from the entire document collection, and subsequently one or more \emph{rerankers} refine the ranking \cite{nogueira2019passage}.
The retriever needs to run fast with high efficiency in order to scan through the entire corpus with low latency. As a result, retrievers have remained simple and give only mediocre performance. With recent deep neural models like BERT~\cite{devlin2018bert} rerankers pushing reranking accuracy to new levels, first-stage retrievers are gradually becoming the bottleneck in modern search engines.

Typical first-stage retrievers adopt a bag-of-words retrieval model that computes the relevance score based on heuristics defined over the \emph{exact word overlap} between queries and documents. Models such as BM25~\cite{bm25} remained state-of-the-art for decades and are still widely used today.  
Though successful, lexical retrieval struggles when matching goes beyond surface forms and fails when query and document mention the same concept using different words~(\emph{vocabulary mismatch}), or share only high-level similarities in topics or language styles. 

An alternative approach for first-stage retrieval is a neural-based, dense embedding retrieval: query words are mapped into a single vector query representation to search against document vectors. Such methods learn an inner product space where retrieval can be done efficiently leveraging recent advances in maximum inner product search (MIPS)~\cite{ShrivastavaL14,FAISS,guo2020accelerating}.  Instead of heuristics, embedding retrieval learns an encoder to understand and encode queries and documents, and the encoded vectors can softly match beyond text surface form.   
However, single vector representations have limited capacity~\cite{attention}, and are unable to produce granular token-level matching signals that are critical to accurate retrieval~\cite{DRMM,salton83}.  

We desire a model that can capture both token-level and semantic-level information for matching. We propose a novel first-stage retrieval model, \textit{Complementary Retrieval Model (\CLEAR)}, that uses dense embedding retrieval to \emph{complement} exact lexical retrieval. 
\CLEAR adopts a single-stage-multi-retriever design consisting of a lexical retrieval model based on BM25 and an embedding retrieval model based on a Siamese framework that uses BERT~\cite{devlin2018bert} to generate query/document embedding representations. Importantly, unlike existing techniques that train embeddings directly for ranking independently~\cite{zamani2018neuralre,chang2020pre}, \CLEAR explicitly trains the embedding retrieval model with a \emph{residual} method: the embedding model is \emph{trained} to build upon the lexical model's exact matching signals and to fix the mistakes made by the lexical model by supplementing semantic level information, effectively learning semantic matching not captured by the lexical model, which we term the un-captured residual.

Our experiments on large-scale retrieval data sets show the substantial and consistent advantages of \CLEAR over state-of-the-art lexical retrieval models, a strong BERT-based embedding-only retrieval model, and a fusion of the two. Furthermore, \CLEAR's initial retrieval provides additive gains to downstream rerankers, improving end-to-end accuracy and efficiency. Our qualitative analysis reveals promising improvements as well as new challenges brought by \CLEAR.



\section{Related Work}

Traditionally, first-stage retrieval has relied on bag-of-words models such as BM25~\cite{bm25} or query likelihood~\cite{lafferty2001document},  
 and has augmented text representations with $n$-grams~\cite{metzler2005markov}, controlled vocabularies~\cite{rajashekar1995combining}, and query expansion~\cite{lavrenko2001relevance}.
Bag-of-words representations can be improved with machine learning techniques, e.g., by employing machine-learned query expansion on bag-of-sparse-features \cite{YaoDC13,ChenD17}, adjusting terms' weights~\cite{dai2019deepct} with BERT~\cite{devlin2018bert}, or adding terms to the document with sequence-to-sequence models~\cite{nogueira2019doc2qry}. However, these approaches still use the lexical retrieval framework and may fail to match at a higher semantic level.  

Neural models excel at semantic matching with the use of dense text representations. Neural models for IR can be
classified into two groups \cite{DRMM}: \emph{interaction-based} and \emph{representation-based} models.  Interaction-based models model interactions between word pairs in queries
and documents. Such approaches are effective for reranking, but are cost-prohibitive for first-stage retrieval as the expensive document-query interactions must be computed online for all ranked documents.

Representation-based models learn a single vector representation for the query or the document and
use a simple scoring function (e.g., cosine or dot product) to measure their relevance.
Representation-based neural retrieval models can be traced back to efforts such as LSI \cite{deerwester1990indexing},  Siamese networks \cite{bromley1993siamese}, and MatchPlus \cite{caid1995learned}. Recent research investigated using modern deep learning techniques to build vector representations: \cite{lee2019latent} and \cite{guu2020realm} used BERT-based retrieval to find passages for QA; \cite{chang2020pre} proposes a set of pre-training tasks for sentence retrieval.
Representation-based models enable low-latency, full-collection retrieval with a dense index. By representing queries and documents with dense vectors, retrieval is reduced to a  maximum inner product search (MIPS)~\cite{ShrivastavaL14} problem. In recent years, there has been increasing effort on accelerating maximum inner product and nearest neighbor search, which led to high-quality implementations of libraries for nearest neighbor search such as hnsw~\cite{malkov2018efficient}, FAISS~\cite{FAISS}, and SCaNN~\cite{guo2020accelerating}. Notably, with these technologies, nearest neighbor search can now scale to millions of candidates with millisecond latency~\cite{FAISS,guo2020accelerating}, and has been successfully used in large-scale retrieval tasks~\cite{lee2019latent,guu2020realm}.  They provide the technical foundation for fast embedding retrieval of our proposed \CLEAR model.

 The effectiveness of representation-based neural retrieval models for standard ad-hoc search is mixed~\cite{DRMM,zamani2018neuralre}. All of the representation-based neural retrieval models share the same limitation --  they use a fixed number of dimensions, which incurs the specificity vs. exhaustiveness trade-off as in all controlled vocabularies \cite{salton83}. 
Most prior research on hybrid models has focused on the reranking stage~\cite{mitra2017learning}. Some very recent research begins to explore hybrid lexical/embedding models. Its focus is mainly on improving the embedding part with weak-supervision~\cite{Kuzi2020LeveragingSA} for low-resource setups, or new neural architectures that use multiple embedding vectors to raise model capacity~\cite{Luan2020SparseDA}.  In these works, embedding models are all trained independently from the lexical models and rely on simple post-training fusion to form a hybrid score. To the best of our knowledge, ours is the first work that investigates jointly training latent embeddings and lexical retrieval for first-stage ad hoc retrieval.


\section{Proposed Method}
\label{sec-model}

\CLEAR consists of a lexical retrieval model and an embedding retrieval model. Between these two models, one's weakness is the other's strength: lexical retrieval performs exact token matching but cannot handle vocabulary mismatch; meanwhile, the embedding retrieval supports semantic matching but loses granular (lexical level) information. 
To ensure that the two types of models work together and fix each other's weakness, we propose a \emph{residual}-based learning framework that teaches the neural embeddings to be complementary to the lexical retrieval.



\subsection{Lexical Retrieval Model}
Lexical retrievers are designed to capture token level matching information. They heuristically combine token overlap information, from which they compute a matching score for query document pairs. Decades of research have produced many lexical algorithms such as vector space models, Okapi BM25~\cite{bm25}, and query likelihood~\cite{lafferty2001document}.  
We use BM25~\cite{bm25} given its popularity in existing systems. 

Given a query $q$ and document $d$, BM25 generates a score based on the  overlapping words statistics between the pair. 
\begin{align}
     & s_\mathrm{lex} (q,d)  = \mathrm{BM25}(q,d) = \sum_{t \in q \cap d} 
     \mathrm{rsj}_t \cdot \frac{\mathrm{tf}_{t,d}}{\mathrm{tf}_{t,d} + k_1 \left\{(1-b) + b \frac{|d|}{l} \right\} } . \label{eq:bm25}
\end{align}
$t$ is a term, $tf_{t,d}$ is $t$'s frequency in document $d$, 
$rsj_t$ is $t$'s Robertson-Sp\"{a}rck Jones weight,
and $l$ is the average document length.
$k_1$ and $b$ are parameters.

\subsection{Embedding Retrieval Model}
\label{subsec:emb-model}
The embedding retrieval model encodes either the query or document text sequence into a dense embedding vector, and matches queries and documents softly by comparing their vector similarity. Generally, the embedding retrieval model can take various neural architectures that encode natural language sequences such as CNN~\cite{textCNN}, or LSTM~\cite{LSTM}, as long as the model outputs can be pooled effectively into a single fixed-length vector for any input. A model capable of deeper text understanding is usually desired to produce high-quality embedding.

This work uses a Transformer \cite{vaswani2017attention} encoder.  We start with pretrained BERT \cite{devlin2018bert} weights and fine-tune the model to encode both queries and documents into vectors in a d-dimension embedding space, i.e., $\mathbf{v}_q, \mathbf{v}_d \in \mathbb{R}^{d}$. The model has a Siamese structure, where the query and document BERT models share parameters $\theta$ in order to reduce training time, memory footprint, and storthe special token $\qry$ to queries and $\doc$ to documents. For a given query or document, the embedding model computes the corresponding query vector $\mathbf{v}_q$ or document vector $\mathbf{v}_d$, following SentenceBERT~\cite{Reimers2019SentenceBERTSE}, by average pooling representations from the encoder's last layers. 
\begin{equation}
    \mathbf{v}_q = \text{AvgPool}[\text{BERT}_\theta(\qry~;~\text{query})]
\end{equation}
\begin{equation}
    \mathbf{v}_d = \text{AvgPool}[\text{BERT}_\theta(\doc~;~\text{document})]
\end{equation}
The embedding matching score $s_{\rm emb} (q,d)$ is the dot product of the two vectors. We use dot product as the similarity metric as it allows us to use MIPS~\cite{FAISS,guo2020accelerating} for efficient first-stage retrieval.
\begin{equation}
s_{\rm emb} (q,d)
= \mathbf{v}_q^{\rm T} ~\mathbf{v}_d \ .
\end{equation}

\subsection{Residual-based Learning}

We propose a novel residual-based learning framework to ensure that the lexical retrieval model and the embedding retrieval model work well together. While BM25 has just two trainable parameters, the embedding model has more flexibility. To make the best use of the embedding model, we must avoid the embedding model ``relearning" signals already captured by the lexical model. Instead, we focus its capacity on semantic level matching missing in the lexical model. 

In general, the neural embedding model training uses hinge loss~\cite{WestonW99} defined over a triplet: a query $q$, a relevant document $d^+$, and an irrelevant document $d^-$ serving as a negative example:
\begin{eqnarray}
    \mathcal{L} = [m - s_{\rm emb}(q, d^+) + s_{\rm emb}(q, d^-)]_+
        \label{eq:loss}
\end{eqnarray}
where $[x]_+ = \max\{0, x\}$, and m is a static loss margin. In order to train embeddings that complement lexical retrieval, we propose two techniques: sampling negative examples $d^-$ from lexical retrieval errors, and replacing static margin $m$ with a variable margin that conditions on the lexical retrieval's residuals.

\paragraph{Error-based Negative Sampling} 
We sample negative examples ($d^-$ in Eq.~\ref{eq:loss}) from those documents mistakenly retrieved by lexical retrieval. Given a positive query-document pair, we uniformly sample irrelevant examples from the top $N$ documents returned by lexical retrieval with probability $p$. 
With such negative samples, the embedding model learns to differentiate relevant documents from confusing ones that are lexically similar to the query but semantically irrelevant. 

\paragraph{Residual-based Margin} Intuitively, different query-document pairs require different levels of extra semantic information for matching on top of exact matching signals. Only when lexical matching fails will the semantic matching signal be necessary. Our negative sampling strategy does not tell the neural model the degree of error made by the lexical retrieval that it needs to fix. To address this challenge, we propose a new residual margin. In particular, in the hinge loss, the conventional static constant margin $m$ is replaced by a linear residual margin function $m_r$, defined over $\text{s}_\text{lex}(q, d^+)$ and $\text{s}_\text{lex}(q, d^-)$, the lexical retrieval scores:
\begin{align}
    m_r(\text{s}_\text{lex}(q, d^+), \text{s}_\text{lex}(q, d^-)) &= \xi - \lambda_\text{train} (\text{s}_\text{lex}(q, d^+) - \text{s}_\text{lex}(q, d^-)),
    \label{eq:mr}
\end{align}
where $\xi$ is a constant non-negative bias term. The difference $\text{s}_\text{lex}(q, d^+)-\text{s}_\text{lex}(q, d^-)$ corresponds to a residual of the lexical retrieval. We use a scaling factor $\lambda_\text{train}$ to adjust the contribution of residual. Consequently, the full loss becomes a function of both lexical and embedding scores computed on the triplet,
\begin{equation}
    \mathcal{L} 
    = [
    m_r(\text{s}_\text{lex}(q, d^+), \text{s}_\text{lex}(q, d^-)) 
    - s_{\rm emb}(q, d^+) + s_{\rm emb}(q, d^-)
    ]_+
\end{equation}
For pairs where the lexical retrieval model already gives an effective document ranking, the residual margin $m_r$ (Eq.~\ref{eq:mr}) becomes small or even becomes negative. In such situations, the neural embedding model makes little gradient update, and it does not need to, as the lexical retrieval model already produces satisfying results. On the other hand, if there is a vocabulary mismatch or topic difference, the lexical model may fail, causing the residual margin to be high and thereby driving the embedding model to accommodate in gradient update.
%
Through the course of training, the neural model learns to encode the semantic patterns that are not captured by text surface forms. When training finishes, the two models will  work together, as \CLEAR.

\subsection{Retrieval with CLEAR}


\CLEAR retrieves from the lexical and embedding index respectively, taking the union of the resulting candidates, and sorts using a final retrieval score: a weighted average of lexical matching and neural embedding scores:
\begin{equation}
    s_\CLEAR(q,d) =\lambda_\text{test}s_\text{lex}(q,d) + s_\text{emb}(q,d) \label{eq:combine}
\end{equation}
We give \CLEAR the flexibility to take different $\lambda_\text{train}$ and $\lambda_\text{test}$ values. Though both are used for interpolating scores from different retrieval models, they have different interpretations. Training $\lambda_\text{train}$ serves as a global control over the residual based margin. On the other hand, testing $\lambda_\text{test}$ controls the contribution from the two retrieval components.
 
\CLEAR achieves low retrieval latency by having each of the two retrieval models adopt optimized search algorithms and data structures.  For the lexical retrieval model, \CLEAR index the entire collection with a typical inverted index. For the embedding retrieval model, \CLEAR pre-computes all document embeddings and indexes them with fast MIPS indexes such as FAISS~\cite{FAISS} or SCANN~\cite{guo2020accelerating}, which
can scale to millions of candidates with millisecond latency.
As a result, \CLEAR can serve as a first-stage, full-collection retriever.

\section{Experimental Methodology}
\label{sec:exp-method}


\paragraph{Dataset and Metrics} 

We use the MS MARCO passage ranking dataset \cite{MSMARCO}, a widely-used ad-hoc retrieval benchmark  with
8.8 millions passages.
The training set contains 0.5 million pairs of queries and relevant passages, where each query on average has one relevant passage\footnote{Dataset is available at \url{https://microsoft.github.io/msmarco/}.}. 
We used two evaluation query sets with different characteristics:

\begin{itemize}
\item\textbf{MS MARCO Dev Queries} is the MS MARCO dataset's official dev set, which has been widely used in prior research~\cite{nogueira2019passage,dai2019deepct}. It has 6,980 queries. Most of the queries have only 1 document judged relevant; the labels are binary.  MRR@10 is used to evaluate the performance on this query set following \cite{MSMARCO}.  We also report the Recall of the top 1,000 retrieved (R@1k), an important metric for first-stage retrieval. 

\item\textbf{TREC2019 DL Queries} is the official evaluation query set used in the TREC 2019 Deep Learning Track shared task \cite{craswell2019overview}. It contains 43 queries that are manually judged by NIST assessors with 4-level relevance labels, allowing us to understand the models' behavior on queries with \emph{multiple, graded relevance judgments} (on average 94 relevant documents per query). NDCG@10, MAP@1k and R@1k are used to evaluate this query set's accuracy, following the shared task.
\end{itemize}

\paragraph{Compared Systems}
We compare \CLEAR retrieval with several first-stage lexical retrieval systems that adopt different techniques such as traditional BM25, deep learning augmented index and/or pseudo relevance feedback.
\begin{itemize}
    \item \textbf{BM25} \cite{bm25}: A widely-used off-the-shelf lexical-based retrieval baseline.
    \item \textbf{DeepCT} \cite{dai2019deepct}: A state-of-the-art first-stage neural retrieval model. It uses BERT to estimate term importance based on context; in turn these context-aware term weights are used by BM25 to replace \textit{tf} in~\autoref{eq:bm25}.
    \item \textbf{BM25+RM3}: RM3 \cite{lavrenko2001relevance} is a popular query expansion technique. It adds related terms to the query to compensate for the vocabulary gap between queries and documents. BM25+RM3 has been proven to be strong \cite{lin2019neural}. 
     \item \textbf{DeepCT+RM3}: \cite{dc20www} shows that using DeepCT term weights with RM3 can further improve upon BM25+RM3. 
\end{itemize}
In addition, we also compare with an embedding only model, \textbf{BERT-Siamese}: This is a BERT-based embedding retrieval model without any explicit lexical matching signals, as described in \autoref{subsec:emb-model}. Note that although BERT embedding retrieval models have been tested on several question-answering tasks \cite{lee2019latent,guu2020realm,chang2020pre}, their effectiveness for ad hoc retrieval remains to be studied. 



\paragraph{Pipeline Systems}
To investigate how the introduction of \CLEAR will affect the final ranking in state-of-the-art pipeline systems, we introduce two pipeline setups.
\begin{itemize}
    \item \textbf{BM25+BERT reranker}: this is a state-of-the-art \emph{pipelined} retrieval system. It uses BM25 for first-stage retrieval, and reranks the top candidates using a BERT reranker \cite{nogueira2019passage}. Both the \textsc{bert-base} and the \textsc{bert-large} reranker provided by \cite{nogueira2019passage} are explored. Note that BERT rerankers use a very deep self-attentive architecture whose computation cost limits its usage to only the reranking stage. 
    
    \item \textbf{\CLEAR\!\!+BERT reranker}: a similar pipelined retrieval system that uses \CLEAR as the first-stage retreiever,  followed by a BERT reranker (\textsc{bert-base} or \textsc{bert-large} reranker from \cite{nogueira2019passage}).
\end{itemize}

\paragraph{Setup}
Lexical retrieval systems, including BM25, BM25+RM3, and deep lexical systems DeepCT and DeepCT+RM3, build upon Anserini \cite{yang2017anserini}. We set $k_1$ and $b$ in BM25 and DeepCT using values recommended by~\cite{dai2019deepct}, which has stronger performance than the default values. The hyper-parameters in RM3 are found through a simple parameter sweep using 2-fold cross-validation in terms of MRR@10 and NDCG@10; the hyper-parameters include the number of feedback documents and the number of feedback terms (both searched over $\{5,10,\cdots,50\}$), and the feedback coefficient (searched over $\{0.1,0.2,\cdots,0.9\}$).
 
Our neural models were built on top of the HuggingFace~\cite{Wolf2019HuggingFacesTS} implementation of BERT. We initialized our models with \textsc{bert-base-uncased}, as our hardware did not allow fine-tuning \textsc{bert-large} models. For training, we use the 0.5M pairs of queries and relevant documents. At each training step, we randomly sample one negative document from the top 1,000 documents retrieved by BM25. We train our neural models for 8 epochs on one RTX 2080 Ti GPU; training more steps did not improve performance. We set $\xi=1$ in Eq.~\ref{eq:mr}. We fixed $\lambda_{\rm train} = 0.1$ in the experiments. For $\lambda_{\rm test}$, we searched over $\{0,1, 0.2, \cdots, 0.9\}$ on $500$ training queries, finding $0.5$ to be the most robust.
Models are trained using the Adam optimizer \cite{KingmaB14} with learning rate $2\times 10^{-5}$, and batch size 28. 
 In pipelined systems, we use BERT rerankers released by Nogueira et al.~\cite{nogueira2019passage}. Statistical significance was tested using the permutation test with $p < 0.05$. 

\section{Results and Discussion}

We study \CLEAR's retrieval effectiveness on a large-scale, supervised retrieval task,  its impact on downstream reranking, and its winning/losing cases.

\subsection{Retrieval Accuracy of CLEAR}

\begin{table*}[t]
\centering
\adjustbox{max width=0.9\linewidth}{
\begin{tabular}{llll lll}
\toprule
\multirow{2}{*}{Type}  & \multirow{2}{*}{Model} & \multicolumn{2}{c}{MS MARCO Dev} & \multicolumn{3}{c}{TREC2019 DL} \\
\cmidrule(lr){3-4} \cmidrule(lr){5-7} & & \small \begin{tabular}[c]{@{}c@{}}  MRR \\ @10   \end{tabular}  & \small R@1k    &  \small \begin{tabular}[c]{@{}c@{}}  NDCG \\ @10   \end{tabular} & \small \small \begin{tabular}[c]{@{}c@{}}  MAP \\
@1k   \end{tabular} & \small R@1k \\ 
\midrule
\multirow{4}{*}{\small Lexical} & {\small 1} BM25 & 0.191$^2$ & 0.864 & 0.506 & 0.377$^5$ & 0.738$^5$ \\
& {\small 2} BM25+RM3 & 0.166 & 0.861 & 0.555$^1$ & 0.452$^{135}$ & 0.789$^{13}$ \\
& {\small 3} DeepCT              & 0.243$^{124}$  & 0.913$^{12}$                                                   & 0.551$^{1}$                                               & 0.422$^{1}$                                                & 0.756$^{1}$                                                  \\
& {\small 4} DeepCT+RM3          &  0.232$^{12}$      & 0.914$^{12}$                                                          & 0.601$^{123}$                                               & 0.481$^{123}$                                                & 0.794$^{13}$                                                  \\ 
\midrule
{\small Embedding} & {\small 5} BERT-Siamese            & 0.308$^{1-4}$  & 0.928$^{123}$                                                  & 0.594$^{123}$                                              & 0.307                                               & 0.584                                                 \\ \midrule
\multirow{3}{*}{\small \begin{tabular}[l]{@{}l@{}}  Lexical+ \\ Embedding   \end{tabular}} &  {\small 6} \CLEAR               & \textbf{0.338}$^{1-5}$  & \textbf{0.969}$^{1-5}$                                                & \textbf{0.699}$^{1-5}$                                               & \textbf{0.511}$^{1-5}$                                                & \textbf{0.812}$^{1-5}$                                                   \\
& ~$-$ w/ Random Sampling &    \ssl{0.241}    & \ssl{0.926}      &   \ssl{0.553}   & \ssl{0.409}    & \ssl{0.779}   \\ 
& ~$-$ w/ Constant Margin$^*$          & \ssl{0.314} & \ssl{0.955}                                            & \ssl{0.664} & \ssl{0.455}                                         & 0.794 \\ 
\bottomrule
\end{tabular}}

\caption{First-stage retrieval effectiveness of \CLEAR on the MS MARCO  dataset, evaluated using two query evaluation sets, with ablation studies.  Superscripts 1--6 indicate statistically significant improvements over methods indexed on the left. $\downarrow$ indicates a number being statistically significantly lower than \CLEAR. \\$^*$: \CLEAR w/ Constant Margin is equivalent to a post-training fusion of BM25 and BERT-Siamese. }
\label{tab:retrieval}
\end{table*}

In this experiment, we compare \CLEAR's retrieval performance with first stage retrieval models described in \autoref{sec:exp-method} and record their performance in \autoref{tab:retrieval}.

\paragraph{\CLEAR vs. Lexical Retrieval} 
\CLEAR outperforms BM25 and BM25+RM3 systems by large margins in both recall-oriented metrics (R@1k and MAP@1k) as well as precision-oriented ones (MRR@10 and NDCG@10). \CLEAR also significantly outperforms DeepCT and DeepCT+RM3, two BERT-augmented lexical retrieval models. DeepCT improves over BM25 by incorporating BERT-based contextualized term weighting, but still use exact term matching.  The results show that lexical retrieval is limited by the strict term matching scheme, showing \CLEAR's advantages of using embeddings for semantic-level soft matching. 


\paragraph{\CLEAR vs. BERT-Siamese Retrieval} 
BERT-Siamese performs retrieval solely relying on dense vector matching. As shown in \autoref{tab:retrieval},  \CLEAR outperforms BERT-Siamese with large margins, indicating that an embedding-only retrieval
is not sufficient. 
Interestingly, though outperforming BM25 by a large margin on MSMARCO Dev queries, BERT-Siamese performs worse than BM25 in terms of MAP@1k and recall on TREC DL queries. The main difference between the two query sets is that TREC DL query has multiple relevant documents with graded relevance levels. It therefore requires a better-structured embedding space to capture this, which proves to be harder to learn here. \CLEAR circumvents this full embedding space learning problem by grounding in the lexical retrieval model while using embedding as complement. 


\paragraph{Ablation Studies}

We hypothesize that \CLEAR's residual-based learning approach can optimize the embedding retrieval to \emph{complement} the lexical retrieval, so that the two parts can generate additive gains when combined. To verify this hypothesis, 
we run ablation studies by \textbf{(1)} replacing the error-based negative samples with random negative samples, and \textbf{(2)} replacing the residual margin in the loss function with a constant margin, which is equivalent to a fusion of BM25 and BERT-Siamese rankings. Using random negative samples leads to a substantial drop in \CLEAR's retrieval accuracy, showing that it is important to train the embeddings on the mistakenly-retrieved documents from lexical retrieval to make the two retrieval models additive. 
Using constant margins instead of residual margins also lowers the performance of the original \CLEAR model. By enforcing a residual margin explicitly, the embedding model is forced to learn to compensate for the lexical retrieval, leading to improved performance. The results confirm that \CLEAR is more effective than a post-training fusion approach where the retrieval models are unaware of each other.


\subsection{Impacts of \CLEAR on Reranking}
\addtolength{\tabcolsep}{2pt}    
\begin{table*}[t]
\centering
\begin{tabular}{l c c c c }
\toprule
\multirow{2}{*}{Retriever} & \multirow{2}{*}{Reranker}  & MSMARCO Dev & TREC DL &Rerank Depth \\ 
& & \small MRR@10 & \small NDCG@10  & $K$\\
\midrule 
{\small $1$} BM25 & - & 0.191 &  0.506 & - \\ 
{\small $2$} \CLEAR & - & 0.338$^{1}$ &  0.699$^{1}$  & - \\ 
\midrule
{\small $3$} BM25 & \textsc{bert-base}  & 0.345$^{1}$ & 0.707$^1$ & 1k \\
{\small $4$} \CLEAR & \textsc{bert-base} & {0.360}$^{123}$ & {0.719}$^{12}$ & 20 \\ 
\midrule
{\small $5$} BM25 & \textsc{bert-large}  & 0.370$^{123}$ & 0.737$^{123}$  & 1k  \\ 
{\small $6$} \CLEAR & \textsc{bert-large} &     {\bf 0.380}$^{1-5}$ & {\bf 0.752}$^{1-5}$ & 100 \\
\bottomrule
\end{tabular}

\caption{Comparing \CLEAR and the state-of-the-art {BM25+BERT Reranker} pipeline on the MS MARCO passage ranking dataset with two evaluation sets (Dev: MS MARCO Dev queries; TREC: TREC2019 DL queries).
We record the most optimal reranking depth for each initial retriever.  Superscripts 1--6 indicate statistically significant improvements over the corresponding methods. }\label{tab:pipeline}
\end{table*}
\addtolength{\tabcolsep}{-2pt}

\begin{figure}[t]
    \centering

    \begin{subfigure}[t]{0.4\textwidth}
    \includegraphics[width=\textwidth]{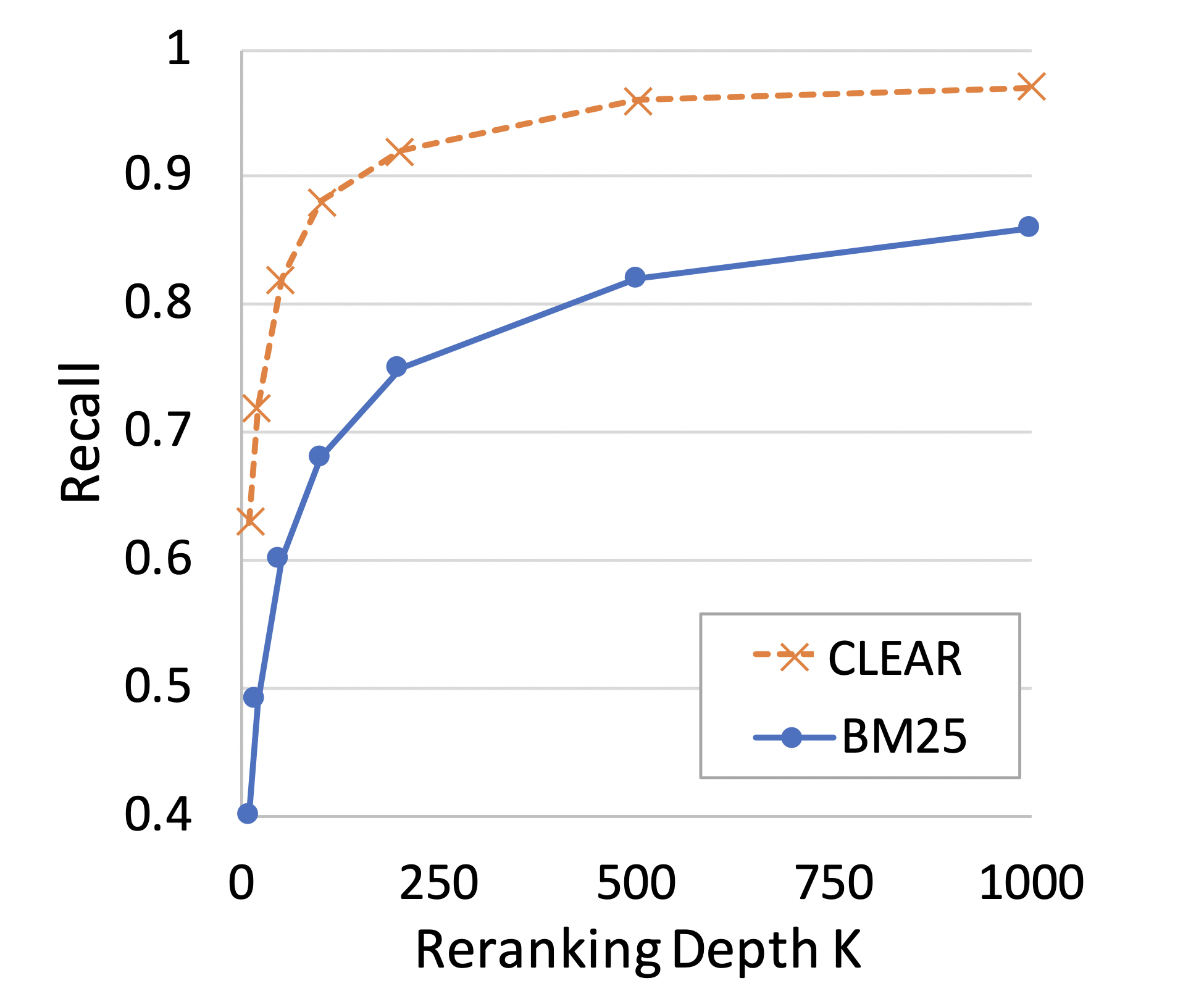}
    \caption{Retrieval Recall}\label{fig:recall}
    \end{subfigure}%
     \begin{subfigure}[t]{0.4\textwidth}
    \includegraphics[width=\textwidth]{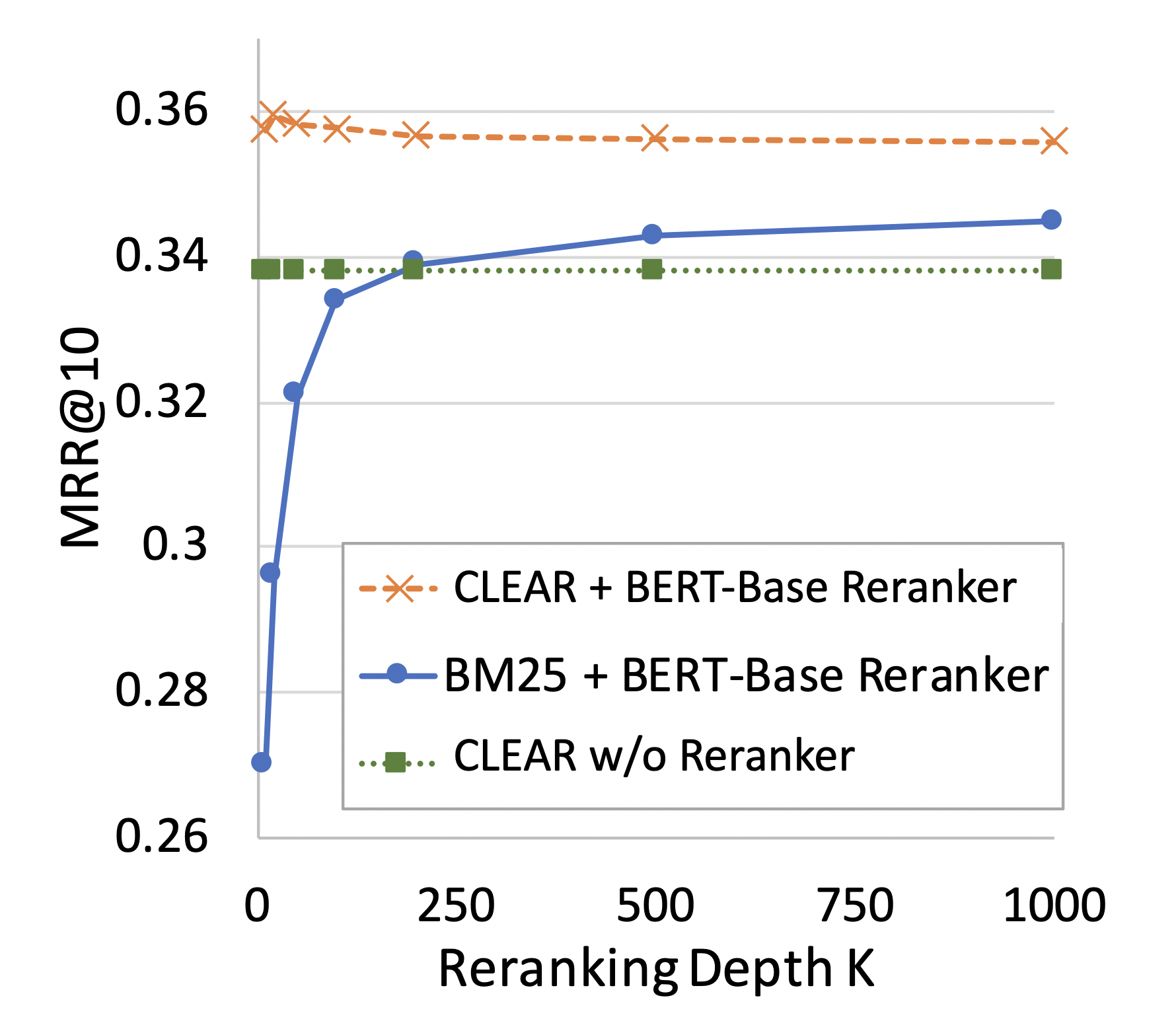}
    \caption{Reranking Accuracy}\label{fig:mrr}
    \end{subfigure}

    \caption{
    Comparison between \CLEAR and BM25 pipeline systems on MS MARCO Dev queries. The system uses the \textsc{bert-base} reranker to rerank against various depth $K$.  }
    \label{fig:rerank}
    \vspace{-4mm}
\end{figure}

Similar to other fist-stage retrievers, \CLEAR can be incorporated into the state-of-the-art pipelined retrieval system, where its candidate list can be reranked by a deep neural reranker. To quantitatively evaluate the benefit of \CLEAR, in the next experiment, we test reranking \CLEAR results with BERT rerankers.

Results are listed in \autoref{tab:pipeline}. Here, we compare \CLEAR against the widely-used BM25 in a two-stage retrieval pipeline, using current state-of-the-art BERT rerankers~\cite{nogueira2019passage} as the second stage reranking model. The rerankers use the concatenated query-document text as input to BERT to classify the relevance. 
We experimented with both \textsc{bert-base} and \textsc{bert-large} reranker variants provided by~\cite{nogueira2019passage}. We also investigate the reranking depth for each initial retriever and record the most optimal here.

The performance of \CLEAR \emph{without reranking} is already close to that of the two-stage BM25+\textsc{bert-base} reranker. 
When adding a reranker, \CLEAR pipelines significantly outperforms the BM25 pipelines. We also discover that reranking a truncated top list for \CLEAR is sufficient, while top 1000 is required for BM25. Concretely, the required re-ranking depth decreased from $K$=1,000 to $K$=20 for \textsc{bert-base} reranker and $K$=100 for \textsc{bert-large} reranker, reducing the computational cost by $10\times$--$50\times$. In other words, \CLEAR generates strong initial rankings that systematically raise the position of relevant documents across all queries and help state-of-the-art rerankers to achieve higher accuracy with lower computational costs, improving end-to-end accuracy, efficiency, and scalability.

\autoref{fig:rerank} further plots the recall and reranking accuracy at various reranking depth. \autoref{fig:recall} shows that \CLEAR had higher recall values than BM25 at all depths, meaning that \CLEAR can provide more relevant passages to the reranker. 
\autoref{fig:mrr} shows the performance of a BERT reranker \cite{nogueira2019passage} applied to the top $K$ documents retrieved from either BM25 or \CLEAR. 
When applied to \texttt{BM25}, the accuracy of the BERT reranker improved as reranking depth K increases.
Interestingly for \CLEAR, the reranking accuracy was already high with small K. While increasing K improves global recall, the reranking accuracy shows saturation with larger K, indicating that BERT rerankers do not fully exploit the lower portion of \CLEAR candidate lists. We investigate this further in \autoref{sec:case-study}.

\subsection{Case Study: The Goods and the new Challenges}
\label{sec:case-study}


\begin{table*}[t]
\centering
\adjustbox{max width=0.98\linewidth}{
\begin{tabular}{l l |c } 

\toprule
Query  & Document retrieved by \CLEAR  & \begin{tabular}[c]{@{}c@{}}  BM25 \\ $\rightarrow$ \CLEAR \end{tabular} \\
\midrule
\textcolor{blue}{weather} in danville, ca
&  \begin{tabular}[l]{p{0.6\textwidth}} 
Thursday:The Danville forecast for Aug 18 is 85 degrees and \textcolor{blue}{Sunny} . There is 24 percentage chance of \textcolor{blue}{rain}  and 10 mph \textcolor{blue}{winds}  from the West.  Friday:...
\end{tabular}
&  \begin{tabular}[c]{@{}c@{}}  989 \\ $\rightarrow$ 10  \end{tabular}                  \\
\midrule brief \textcolor{blue}{government} definition 
&  \begin{tabular}[l]{p{0.6\textwidth}} Legal Definition of brief. 1  1 : a concise statement of a client's case written for the instruction of an \textcolor{blue}{attorney} usually by a \textcolor{blue}{law clerk} ...
\end{tabular}  
& \begin{tabular}[c]{@{}c@{}} 996 \\ $\rightarrow$ 7  \end{tabular} \\
\midrule  
population of \textcolor{blue}{jabodatek} 
& \begin{tabular}[l]{p{0.6\textwidth}}  The population of \textcolor{blue}{Jabodetabek}, with an area of 6,392 km2, was over 28.0 million according to the Indonesian Census 2010 ....
\end{tabular}  
&  \begin{tabular}[c]{@{}c@{}} not retrieved \\ $\rightarrow$ 1  \end{tabular}\\
\bottomrule

\end{tabular}}
\caption{Example documents retrieved by \CLEAR. We show ranking improvements from pure BM25 to \CLEAR's complementary setup .}\label{tab:improved-cases}
\vspace{-3mm}
\end{table*}

\begin{table*}[t]
\centering
\adjustbox{max width=0.98\linewidth}{
\begin{tabular}{l l | c} 

\toprule
Query  & Document retrieved by \CLEAR & \begin{tabular}[c]{@{}c@{}}\CLEAR\\ $\rightarrow$ Rerank \end{tabular} \\
\midrule
who is robert \textcolor{red}{gray} & \begin{tabular}[l]{p{0.6\textwidth}} \textcolor{red}{Grey} started ... dropping his Robert Gotobed alias and using his birthname Robert \textcolor{red}{Grey}.  \end{tabular} 
&  \begin{tabular}[c]{@{}c@{}} rank 496\\ $\rightarrow$ rank 7 \end{tabular}                  \\
\midrule
\begin{tabular}[c]{@{}l@{}}what is \textcolor{red}{theraderm}  \\used for \end{tabular}         
& \begin{tabular}[l]{p{0.6\textwidth}}A \textcolor{red}{thermogram} is a device which measures heat through use of picture ....\end{tabular} 
&  \begin{tabular}[c]{@{}c@{}} rank 970\\ $\rightarrow$ rank 8 \end{tabular}\\
\midrule
\begin{tabular}[c]{@{}l@{}}what is the daily life \\of \textcolor{red}{thai people} \end{tabular}  
& \begin{tabular}[l]{p{0.6\textwidth}}Activities of daily living include are the tasks that are required to get going in the morning ... 1 walking. 2  bathing.  3 dressing. \end{tabular}
&  \begin{tabular}[c]{@{}c@{}}rank 515\\ $\rightarrow$ rank 7 \end{tabular}\\
\bottomrule
\end{tabular}}
\caption{Challenging non-relevant documents retrieved only by \CLEAR, not by BM25, through semantic matching. We show in \CLEAR initial candidate list ranking as well as after BERT reranking.}
\label{tab:confusing-cases}
\vspace{-5mm}
\end{table*}

In this section, we take a more in-depth look into \CLEAR through case studies. We first examine how BM25 ranking changes after being complemented by the dense embedding retrieval in \CLEAR, then turn to investigate why the lower part of \CLEAR's candidates are challenging for BERT rerankers.

In \autoref{tab:improved-cases}, we show three example queries to which the \CLEAR brings huge retrieval performance improvement. We see that in all three queries, critical query terms, \textit{weather}, \textit{government} and \textit{jabodatek}, have no exact match in the relevant document, leading to failures in exact match only BM25 system. \CLEAR solves this problem, complementing exact matching with high-level semantic matching. As a result, ``weather" can match with document content ``sunny, rain, wind" and ``government" with document content ``attorny, law clerk". In the third query, spelling mismatch between query term ``jabodatek" and document term ``Jabodetabek" is also handled. 

While \CLEAR improves relevant documents' rankings in the candidate list, it also 
brings in new forms of non-relevant documents that are not retrieved by lexical retrievers like BM25, and affects downstream rerankers. %
In \autoref{tab:confusing-cases}, we show three queries and three corresponding false positive documents retrieved by \CLEAR, which are not retrieved by BM25.   Unlike in BM25, where false positives mostly share surface text similarity with the query, in the case of \CLEAR, the false positives can be documents that are topically related but not relevant. In the first two queries, \CLEAR mistakenly performs soft spell matches, while in the third one critical concept ``thai people" is ignored.  

Such retrieval mistakes further affect the performance of downstream BERT reranker. As BERT also performs semantic level matching without explicit exact token matching to ground, the rerankers can amplify such semantically related only mistakes. 
As can be seen in \autoref{tab:confusing-cases},
those false positive documents reside in the middle or at the bottom of the full candidate list of \CLEAR. With BERT reranker, however, their rankings go to the top.  
In general, \CLEAR goes beyond exact lexical matching to rely on semantic level matching. While improving initial retrieval, it also inevitably brings in semantically related false positives. Such false positives are inherently more challenging for state-of-the-art neural reranker and require more robust and discriminative rerankers.  We believe this also creates new challenges for future research to improve neural rerankers.
\section{Conclusion}

Classic lexical retrieval models struggle to understand the underlying meanings of queries and documents. Neural embedding based retrieval models can soft match queries and documents, but they lose specific word-level matching information. 
This paper present \CLEAR, a retrieval model that complements lexical retrieval with embedding retrieval. Importantly, instead of a linear interpolation of two models, the embedding retrieval in \CLEAR is exactly trained to fix the errors of lexical retrieval.

Experiments show that \CLEAR achieves the new state-of-the-art first-stage retrieval effectiveness on two distinct evaluation sets, outperforming classic bag-of-words, recent deep lexical retrieval models, and a BERT-based pure neural retrieval model.  The superior performance of \CLEAR indicates that it is beneficial to use the lexical retrieval model to capture simple relevant patterns using exact lexical clues, and complement it with the more complex semantic soft matching patterns learned in the embeddings. 


Our ablation study demonstrates the effectiveness of 
 \CLEAR's residual-based learning. The error-based negative sampling allows the embedding model to be aware of the mistakes of the lexical retrieval, and the residual margin further let the embeddings focus on the harder errors. Consequently, \CLEAR outperforms post-training fusion models that directly interpolate independent lexical and embedding retrieval models' results.

A single-stage retrieval with \CLEAR achieves an accuracy that is close to popular two-stage pipelines that uses a deep Transformer BERT reranker. We view this as an encouraging step towards building deep and efficient retrieval systems. When combined with BERT rerankers in the retrieval pipeline, \CLEAR's strong retrieval performance leads to better end-to-end ranking accuracy and efficiency. However, we observe that state-of-the-art BERT neural rerankers do not fully exploit the retrieval results of \CLEAR, pointing out future research directions to build more discriminative and robust neural rerankers.  
\bibliographystyle{splncs04}
\bibliography{emnlp-ijcnlp-2019.bib}



\end{document}